# Measuring height difference using two-way satellite time and frequency transfer

Peng Cheng[1], WenBin Shen[1,2*], Xiao Sun[1], Chenghui Cai[1], Kuangchao Wu[1], and Ziyu Shen[3]

[1] Time and Frequency Geodesy Center, School of Geodesy and Geomatics/Key Laboratory of Geospace Environment and Geodesy of Ministry of Education, Wuhan University, Wuhan, China

[2] State Key Laboratory of Information Engineering in Surveying, Mapping and Remote Sensing, Wuhan University, Wuhan, China.

[3] School of Resource, Environmental Science and Engineering, Hubei University of Science and Technology, Xianning 437100, China

**Abstract:** According to general relativity theory (GRT), the clock at a position with lower geopotential ticks slower than an identical one at a position with higher geopotential. Here, we provide a geopotential comparison using a non-transportable hydrogen clock and a transportable hydrogen clock for altitude transmission based on the two-way satellite time and frequency transfer (TWSTFT) technique. First, we set one hydrogen clock on the fifth floor and another hydrogen clock on the ground floor, with their height difference of 22.8 m measured by tape, and compared the time difference between these two clocks by TWSTFT for 13 days. Then, we set both clocks on the ground floor and compared the time difference between the two clocks for 7 days for zero-baseline calibration (synchronization). Based on the measured time difference between the two clocks at different floors, we obtained the height difference 28.0±5.4 m, which coincides well with the tape-measured result. This experiment provides a method of height propagation using precise clocks based on the TWSTFT technique.

1. Introduction

The gravity potential (geopotential) plays a significant role in geodesy. It is essential for defining the geoid and measuring orthometric height. The conventional method of determining the geopotential is combining leveling and gravimetry, but there are shortcomings: with the increase in measurement length, the error accumulates and becomes larger and larger , and it is impossible or difficult to transfer the orthometric height between two points separated by oceans.

To overcome the shortcomings existing in the conventional method, time–frequency comparison methods based on general relativity theory (GRT) [1] were proposed in recent decades [2-7]. The basic idea is that by comparing the time elapsed or frequency shift between two remote clocks, the geopotential difference between the two sites where the clocks are located could be determined.

In the clock-transportation method, the most critical conditions are the clock's precision and time transfer. Precise clocks generally include microwave-atomic clocks (MACs) and optical-atomic clocks (OACs). MACs play an important role in time service research. However, with the development of a high-precision clock, its precision could not match that of the OAC. The concept of OAC was first proposed by Nobel laureate Dehmelt (1973) in the late 1970s. He used the energy level transition of a single ion to realize an ultra-high-precision optical clock. In recent years, the precision of OAC has reached a total systematic uncertainty of 9.4×10$^{-19}$ and frequency stability of 1.2×10$^{-15}$ /$\sqrt{\tau}$ [8]. However, OACs are often bulky and can only work in laboratory environments, which significantly limits their application scope and makes it difficult to conduct a clock-transportation experiment. It has always been the wish of scientists to realize a transportable, reliable, and quasi-continuous high-precision OAC, but it is also challenging. To broaden the application scope of OAC, many research groups in the worldwide have been devoted to the development of transportable optical-atomic clocks (TOCs). In 2014, a group reported a TOC based on laser-cooled strontium atoms trapped in an optical lattice, and this TOC fits within a volume of <2 m$^3$, and its relative uncertainty is 7×10$^{-15}$ [9]. Three years later, a group from Physikalisch-Technische Bundesanstalt (PTB) reported a TOC with $^{87}$Sr, with its characterization against an OAC resulting in a systematic uncertainty of 7.4 × 10$^{-17}$ [10]. With the development of TOC, many scientists started to conduct clock-transportation experiments. Grotti et al. (2018) reported the first field measurement campaign with a $^{87}$Sr TOC with an uncertainty of 1.8×10$^{-16}$ and a $^{171}$Yb OAC with an uncertainty of 1.6×10$^{-16}$. They used these clocks and fiber link to determine the geopotential difference between the middle of a mountain and a location 90 km away with a height difference of 1000 m. Their experimental result of potential difference of 10,034(174) m$^2$s$^{-2}$ agrees well with value of 10,032.1(16) m$^2$s$^{-2}$ determined independently by the conventional geodetic approach [3]. Takamoto et al. (2020) conducted the experiment of gravitational redshift observed by optical fiber connecting two optical clocks. A pair of S$_r$ TOC with the uncertainty of 5×10$^{-18}$ were placed on the tower of the Tokyo Skytree and the ground, respectively (height difference 450 m). Finally, they obtained a height difference 452.631±0.013 m, which agrees well with the value of 452.652±0.039 m measured by GNSS, leveling, and laser ranging [2].

The clock-transportation experiments mentioned above used optical fiber to transfer frequency signals. Although optical fiber has very high accuracy, distance still limits its application. By comparison, though GNSS common-view technique and TWSTFT have lower accuracy than optical fiber frequency signal transfer [11-15], they can realize long-distance time–frequency signal transmission. In TWSTFT, one uses a geostationary satellite as 'bridge' to transmit time–frequency signals from one station to another one, and the time elapse recorded at one station is compared with that at another one. Because using TWSTFT only requires the stations in a place where the geostationary satellite can receive and transmit signals, there is hardly any limit on the positions of stations. Most errors in the signal-transmission process are offset because of the approximate symmetry of the signal transmission path of TWSTFT technology. Therefore, this symmetry causes the high precision of this technology [16]. In August 1962, the USNO (U.S. Naval Observatory) launched the first communication satellite, Telstar I, to transmit telephone and high-speed data signals. Then, the USNO and the NPL (National Physical Laboratory) collaborated in an experiment using this satellite to relate the precise clocks at the USNO and the RGO (Royal Greenwich Observatory). This is considered to be the first two-way satellite time transfer experiment, and the accuracy of experiment is 20 $\mu s$ [17]. In 1992, some satellite systems and modems adapted for TWSTFT were commercialized. About ten coordinated universal time (UTC) laboratories have equipped with the modems and other equipments for the clock comparisons with TWSTFT [18]. Later, many TWSTFT experiments were conducted, and they almost exchanged timing information via the communication satellite; paired ground stations transmit and receive pseudo-random noise (PN) coded signals in TWSTFT links. The TWSTFT technique became promising using geostationary satellites for high-accuracy time and frequency transfer [16, 19-21]. The accuracy of TWSTFT has been further improved to around 0.2 ns, and its improvement has been seriously limited by the chip rate of the coded signal [22]. Therefore, further improvements should come from the use of carrier phase information, because the resolution of the carrier phase is 100 to 1000 times more accurate than that of the code [23]. In 2016, an experiment using carried-phase TWSTFT was performed between the two stations of NICT (National Institute of Information and Communications Technology) and KRISS (Korea Research Institute of Standards and Science) with Sr and Yb OAC, and the instability for a frequency transfer at the $10^{-16}$ level after 12 hours was achieved [24]. Riedel et al. (2020) conducted a 26-day comparison of five simultaneously operated OACs and six MACs located at SYRTE, NPL, INRIM, LNE, and PTB by using TWSTFT and GPSPPP. Considering the correlations and gaps of measurement data, they improved the statistical analysis procedure; combined overall uncertainties in the range of $1.8 \times 10^{-16}$ to $3.5 \times 10^{-16}$ for the OAC comparisons were found[25]. To investigate the feasibility of transportable atomic clock comparison using TWSTFT, we conducted a MAC comparison experiment at the Beijing Institute of Radio Metrology and Measurement (BIRMM), Beijing [26, 27].

In Section 2, we introduce the principle of measuring the height difference by the time difference, and in Section 3, we describe the time-transfer process between two clocks based on TWSTFT. Section 4 discusses the error corrections of TWSTFT. Section 5 demonstrates our experiment and data processing. In section 6 provides the results. Conclusions and discussions are placed in Section 7.

## 2 Methods

*2.1 Determination of time–frequency difference between two clocks based on TWSTFT*

According to general relativity theory (GRT), a clock at a position with lower geopotential ticks slower than an identical at a position with higher geopotential [5, 28]. Inversely, we can determine the geopotential difference between two points A and B by measuring the elapsed time difference between two clocks located at A and B, expressed as (accurate to $1/c^2$) [7, 29, 30]

$$\frac{\Delta t_{AB}}{T} = \frac{t_B - t_A}{T} = \frac{W_B - W_A}{c^2} = \frac{\Delta W_{AB}}{c^2} \tag{1}$$

where $t_A$ and $t_B$ denote the times at sites *A* and *B*, respectively, after a standard time period of T; $W_A$ and $W_B$ are the geopotential at sites *A* and *B*, respectively; and c is the speed of light in the vacuum. From Eq. (1), we can define the geopotential difference $\Delta W_{AB} = W_B - W_A$ based on $\Delta t_{AB}/T$.

In another aspect, based on the gravitational redshift effect, if we may determine the gravity frequency shift (the frequency difference caused by the gravity field) between A and B, we may also determine the geopotential difference $\Delta W_{AB}$, expressed as [6, 28, 31]

$$\frac{\Delta f_{AB}}{f} = \frac{f' - f}{f} = \frac{\Delta W_{AB}}{c^2} \tag{2}$$

where $f$ is the emitting frequency at A and $f'$ is the receiving frequency at B.

After the time difference $\Delta t_{AB}$ or frequency difference $\Delta f_{AB}$ is determined, the geopotential difference $\Delta W_{AB}$ can be determined.

Given the height of point A and the geopotential difference $\Delta W_{AB}$ between A and B, one can measure the orthometric height of point B (Fig. 1), expressed as [32, 33]

$$H_B = H_A + \frac{\Delta W_{AB}}{\bar{g}} \quad (3)$$

where $H_A$ and $H_B$ are the orthometric heights of point A and B, respectively, and $\bar{g}$ is the 'mean value' between $g_B$ at point B and $g_{O_B}$ at point $O_B$.

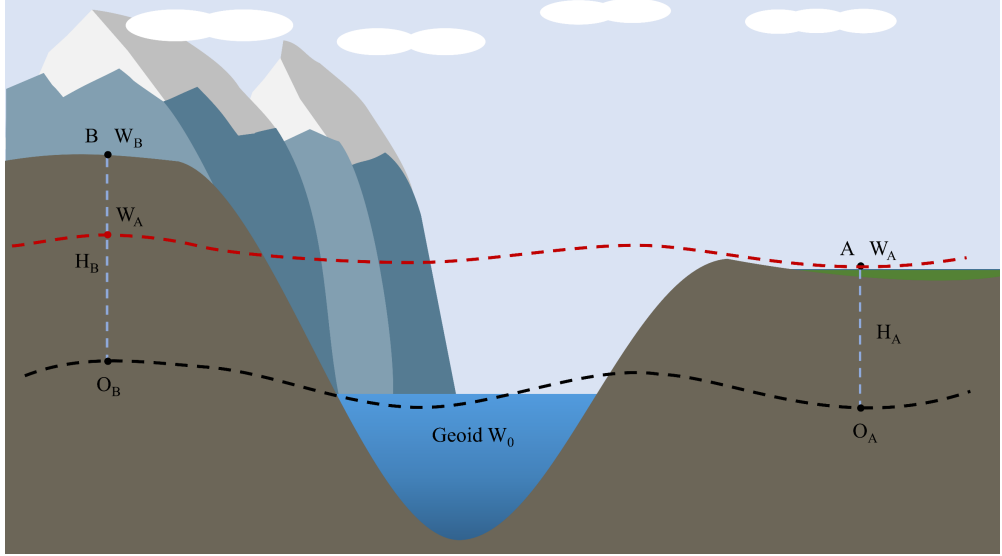

**Figure 1**. Principle of determining the orthometric heights of point B via the orthometric heights of point A. $W_A$ and $W_B$ is the geopotential of point A and point B, respectively; $O_A$ and $O_B$ are the projection points on the geoid of point A and point B along the plumb lines (denoted by blue dashed curves), respectively. Black horizontal dashed curve denotes geoid, and the geopotential of the points above it is $W_0$. Red horizontal dashed curve denotes equipotential surface passing point A.

*2.2 One pulse per second (1 PPS) signal*

The 1 PPS signal provides precise clock synchronization [34]. The original signal is the frequency signal (with typical frequency 10 MHz) output by the atomic clock. As an analog signal, in distant transmission, the information carried by the original frequency signal is easily distorted by various interferences [35]. Therefore, it needs to be converted into a digital signal, 1PPS, to complete the time-information transmission. The process of converting frequency signal (say 10 MHz) into 1PPS signal is shown in Fig. 2. When the 1PPS signal is used for time signal comparison, the rising edge of the signal will be used as the point to trigger the timer. Usually, the electrical level of the trigger is a predetermined value between the lowest electrical level and the highest electrical level. Ideally, when the electrical level rises to this predetermined value, the switch is triggered immediately, and the timing starts. However, there is a trigger delay during the process of the trigger switch. In fact, the switch can only be triggered when the actual electrical level is slightly higher than the predetermined value, so the trigger time will be within the time corresponding to the red oblique line [36]. Hence, the rising time duration $\delta t$ is very important for precise time synchronization. Usually, $\delta t$ should be smaller than 10 ns, since if $\delta t$ is too large, the rising edge's slope (the blue slope line at the bottom of Fig. 2) will become too small, which will increase the uncontrollable duration (red line) and reduce the time synchronization precision.

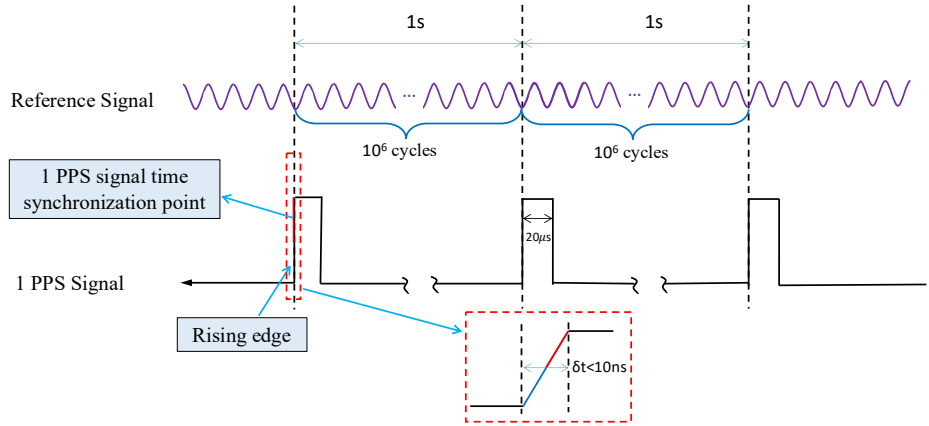

**Figure 2**. Generation of the one pulse per second (1 PPS) signal. Suppose the frequency of the reference signal is 10 MHz; therefore, $10^6$ cycles of the reference signal is one second. The generated 1 PPS signal rises from a low electrical level at the beginning of one reference signal cycle, and keeps the high electrical level for a short time (generally the pulse width is 20 μs), then declines to a low electrical level and keeps the site until the end of the $10^6$ cycles (calculated from the first rise). This is a complete cycle9 (1 s) of 1 PPS signal.

*2.3 Transmission of 1 PPP signal*

If the 1PPS signal is directly transmitted to another station, it is difficult to compare the 1PPS signals generated by two clocks at two sites synchronously. Hence, using the PN code to accompany the 1PPS signal for time marking is necessary. The maximal length linear feedback shift register sequences (m-sequences) is the PN code usually used in TWSTFT. The m-sequence is the largest code that can be generated by a given shift register or a delay element of a given length [37]. The code has good autocorrelation characteristics and can be easily generated (Enge et al. 1987). At the same time, to reduce the impact of the bit error on the final time comparison, the frame synchronization bits (FSB) will be inserted at the rising edge of the 1PPS signal. The bit error means that in signal transmission, decay causes the signal to be damaged, so the originally transmitted signal is '1', but the received signal is '0' (it should also be '1'), and the FSB is a specific set of bits at the beginning of each frame of signal [38]. The spreading spectrum consists of 1PPS and PN, which is completed when the 1PPS signal combines with the synchronized PN code by way of modulo-2 sum in the code generator (CG). This spread spectrum signal (SSS) is the baseband signal with wide bandwidth and decays rapidly during transmission. If it is directly used for long-distance baseband transmission, due to quick attenuation, the signal at the receiver will have a too-low signal-to-noise ratio (SNR) to be identified [39]. Therefore, the SSS needs to be modulated on a proper carrier. Generally, the carrier is a cosine wave, with its amplitude, frequency, and phase being known, and is used to transmit the information (signal) by changing the carrier's amplitude, or frequency, or phase, or a combinations of these entities.

The SSS is modulated on the carrier in the modulator with the binary phase-shift keying (BPSK) [40]. The BPSK changes the initial phase of the carrier to transmit the binary digital information of the SSS, while the carrier's amplitude and frequency remain unchanged (in the modulation process). In this modulation technology, the initial phase of the carrier has only two values of '0' and 'π', which correspond to the digits '1' and '0' of the 1PPS signal, respectively. When transmitting signal '1', the modem generates a carrier with initial phase 0, and when transmitting signal '0', the modem generates a carrier with the initial phase π. The modulated signal is an intermediate frequency (IF) signal, which is named 1PPSTX.

Referring to Fig. 3, the frequency of the 1PPSTX signal generated at station A is about 70 MHz, which is not suitable for long-distance signal transmission. Therefore, the frequency of the 1PPSTX signal is amplified to 14 GHz through the multiplier in the up-converter (UC) and transmitted to the geostationary satellite through the transmitter.

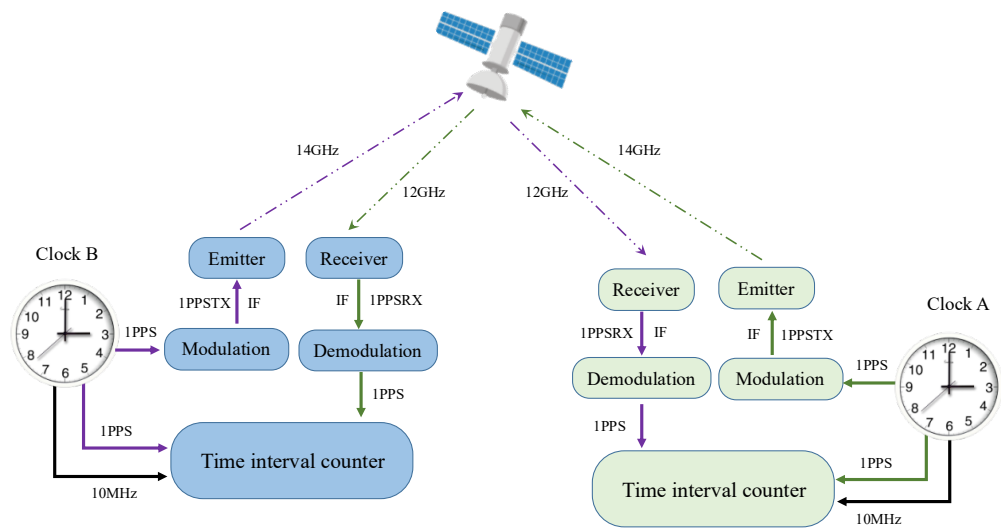

**Figure 3**. Principle of the TWSTFT. The local clock generates the 1PPS signal and the 10 MHz signal output. The 10 MHz signal and part of the 1PPS signal enter time interval counter (TIC) as a timing reference and trigger gate open pulse, respectively. Another part of the 1PPS signal is converted into an IF signal (70 MHz), 1PPSTX, through a modulation module. Then, 1PPSTX is converted into a 14 GHz signal through the signal-transmission module and transmitted to the satellite. When receiving a signal from a ground station, the satellite uses a transparent transponder to convert the signal's frequency to 12 GHz and transmits it to another ground station. After the signal from the satellite is received, the receiving module converts the 12 GHz signal into the IF signal (70 MHz), 1PPSRX. 1PPSRX enters the demodulation module, recovers to the 1PPS signal, and then enters the TIC as a pulse to trigger the gate close. The above processes are conducted between each of the two stations (modified after ITU-R 2015))

After it receives the up-link signal, the satellite transparent transponder (STT) converts the signal's frequency to 12 GHz and transmits the signal (as a down-link signal) to station B. When the down-link signal arrives at station B, the signal's frequency is reduced to about 70 MHz through the multiplier in the down-converter (DC) (the function of which is similar to UC, but in the opposite way). At the same time, the received signal will also be digitized to facilitate subsequent processing. Usually, when this signal is converted to the IF range, a frequency shift will impact the signal acquisition. The reason is that the carrier used to demodulate the IF signal to the baseband signal needs to have the same frequency as the IF signal. When capturing the received signal, the synchronization information of the signal is searched, and the known PN code is used to synchronize the received signal and PN code. Then, the local carrier frequency and code phase are almost identical to the received signal. After the capture is implemented, the signal will be tracked in the delay-locked loop (DDL), which can fully synchronize the local carrier and the PN code with the received signal, thus removing the carrier and PN code from the received signal and recovering the 1PPS signal. Thus far, we have completed the signal demodulating and despreading and obtained the transmitting 1PPS signal generated by station A [41]. However, due to the existence of bit error, it is necessary to detect the synchronization code in the 1PPS signal to ensure that the correct synchronization point of the 1PPS signal is generated [38]. Since it has been synchronized by PN code, the recovered 1PPS signal needs to be input to TIC to compare with the local 1PPS signal. Each rising edge of the local 1PPS signal will be used as the opening point of the timing gate, and each corresponding rising edge of the recovered 1PPS signal will be used as the closing point of the timing gate. In the TIC, the 10 MHz signal output by the local clock is used as the time base reference, measuring the time length between the opening and the closing each time, and then the TIC outputs the time difference as the observed value. Therefore, at both clock sites, the time signals are transmitted nominally at the same instant. Each clock site receives the signal from the other clock site, and its arrival time is measured. After exchanging the measured data, the time–frequency difference between the two clocks is calculated.

**3 Two-way satellite time and frequency transfer (TWSTFT)**

The TWSTFT is based on the exchange of timing signals through geostationary telecommunication satellites. It is completed by transmission and reception of the IF signal, which contains PN codes with the BPSK modulated on the intermediate frequency (IF) by a modem. The main ground station for TWSTFT is equipped with atomic clocks, modems, a very small aperture terminal (VSAT), and TIC. The main instruments equipped on communications satellites are transparent transponders, which amplify the received signal and convert the up-link frequency to the down-link frequency. The TWSTFT exchanges signals between two stations as follows.

Station A generates a 10 MHz frequency signal by the clock; then, the signal is transmitted to the modem to generate a 1 PPS signal. Part of the 1 PPS synchronous output becomes the signal called 1 PPSTX with a BPSK sequence. The 1 PPSTX signal is used to record the signal's transmitting time in the TIC. Another part is modulated to an intermediate frequency (IF) signal of 70 Hz by the modem; then, through signal frequency amplifier, it becomes an up-link frequency signal (usually 14 GHz) transmitted by the transmitter. The satellite receives the signal from station A, converts the up-link frequency signal into the down-link frequency signal (usually 12 GHz) through the transponder by the technique of frequency-mixing [23], and forwards it to station B. Station B (receiver) receives the down-link signal, the receiving equipment generates the BPSK sequence of Station A (transmitter), and reconstitutes a 1PPS signal from the received signal, named 1PPSRX. The TIC measures the difference $\tau_I^k$ between the two 1PPS signals at station B. In the same way and procedures, station B transmits and station A receives signals. According to the schedule, the two stations lock the coding of the corresponding remote radio station in a specific period of time, called a session, and measure the arrival time of the signal and store the results. After exchanging data records, the difference between the two clocks can be calculated[42].

The TIC reading at station A is expressed as:
$$\tau_I^A = \tau^A - \tau^B + \tau_t^B + \tau_{pu}^B + \tau_{su}^B + \tau_s^B + \tau_{pd}^A + \tau_{sd}^A + \tau_r^A \tag{4}$$

and that at station B is expressed as:
$$\tau_I^B = \tau^B - \tau^A + \tau_t^A + \tau_{pu}^A + \tau_{su}^A + \tau_s^A + \tau_{pd}^B + \tau_{sd}^B + \tau_r^B \qquad (5)$$
Combining Expressions (4) and (5), the time scale difference could be expressed as:
$$\tau^A - \tau^B = \frac{1}{2}(\tau_I^A - \tau_I^B) + \frac{1}{2}(\tau_s^A - \tau_s^B) - \frac{1}{2}(\tau_{sd}^A - \tau_{su}^A) + \frac{1}{2}(\tau_{sd}^B - \tau_{su}^B)$$
$$+ \frac{1}{2}(\tau_{pu}^A - \tau_{pd}^A) - \frac{1}{2}(\tau_{pu}^B - \tau_{pd}^B) + \frac{1}{2}(\tau_t^A - \tau_r^A) - \frac{1}{2}(\tau_t^B - \tau_r^B) \qquad (6)$$

where $\tau^k$ denotes the local time-scale, and k means station k (k = A, B); $\tau_I^k$ is the time interval reading; $\tau_{su}^k$ and $\tau_{sd}^k$ are the Sagnac effect delay in the up-link and down-link, respectively; $\tau_{pu}^k$ and $\tau_{pd}^k$ the signal path up-link and down-link delays, respectively; $\tau_s^k$ is the satellite path delay through the transponder, $\tau_t^k$ the transmitter delay, and $\tau_r^k$ the receiver delay [42].

**4 Error analysis and corrections**

The error sources in TWSTFT observations mainly come from three aspects: (1) equipment delay errors; (2) signal-propagation-path-delay errors; and (3) Sagnac effect errors.

*4.1 Equipment delay error*

Equipment delay error mainly includes TIC measurement error, modem errors, satellite transparent transponder delay error, and the emitting and receiving delay error of Earth's station equipment. The TIC measurement error and modems error, which are caused by their own measurement accuracy errors, are about dozens of picoseconds and 100 ps [42], respectively. The satellite transparent transponder delay error, mainly due to the signal from A to B and the signal from B to A through different forwarding channels of the transparent transponder, is hard to control, and is generally in the range of 100 ps [43]. The emitting and receiving equipment delay error of the ground–station is around 0.2~0.5 ns, including cable delay, the transmission and receiving system error, and temperature variation.

It may not be accurate enough to measure the delay error of the signal after passing through each piece of equipment alone. Therefore, the zero-baseline measurement could be a better way to measure all equipment-delay errors. Zero-baseline measurement is when two atomic clocks are placed at the same place simultaneously, and then a TWSTFT experiment is conducted.

*4.2 Propagation path delay error*

The signal's propagation path delay error is caused by the satellite's signal path delay errors of up-link and down-link, mainly including the delay propagation path geometry error and tropospheric delay error, ionospheric delay error, and delay errors of the different distances between two stations and the satellite.

The propagation path geometry delay is related to the coordinates of satellites and ground stations. For the signal's arriving and receiving time, if there is an error for ground stations or satellite position, the error will directly affect the time delay between the satellite and ground station. However, the clock difference calculation formula includes the difference between the up-link $\tau_{pu}^k$ and down-link path $\tau_{pd}^k$; therefore, through the difference between two paths, one can eliminate the error with only a few picoseconds left. That means, at the same time, coordinate errors of satellite and ground stations do not affect the calculation of the clock difference.

Tropospheric delay is also called the tropospheric refraction error. The troposphere, through changing the propagation path of the signal, causes time delay. Many factors can affect tropospheric delay, such as ground climate, atmospheric pressure, temperature, and humidity TEC. Tropospheric delay can use the tropospheric delay model to correct; common models are the Hopfield model, Saastamoinen model, EGNOS model, etc. The differences between each model are mainly in the low-altitude angle; at the zenith direction, the difference is very small. Moreover, the up-link and down-link paths are symmetrical, hence, the tropospheric delay can be mostly cancelled. The remaining time delay is less than 10ps, and it can be neglected in the present study [42].

The ionospheric delay error is mainly caused by charged particles in the ionosphere. The charged particles can change the speed and path of the signal. The ionospheric delay depends on the total electron content (TEC) and the signal's frequency. Because the up-link and down-link signals at each station differ in carrier frequency, the following formula is used to correct the ionospheric delay error [44]:

$$\tau_{pu}^{k}\big|_{\text{ion}} - \tau_{pd}^{k}\big|_{\text{ion}} = \frac{40.28\, e_{ks}}{c}\left(\frac{1}{f_U^2} - \frac{1}{f_D^2}\right) \tag{7}$$

where $f_U$ and $f_D$ are the up-link and down-link frequencies of the signals, respectively; $e_{ks}$ is the TEC along the signal-propagation path between ground station k and satellite S, and c is the speed of light in vacuum. Thus, we have the following equation to correct the ionospheric delay:

$$\frac{1}{2}\left[\tau_{pu}^{A}\big|_{ion} - \tau_{pd}^{A}\big|_{ion}\right] - \frac{1}{2}\left[\tau_{pu}^{B}\big|_{ion} - \tau_{pd}^{B}\big|_{ion}\right] = \frac{20.14(e_{As}-e_{Bs})}{c}\left(\frac{1}{f_U^2} - \frac{1}{f_D^2}\right) \tag{8}$$

Obviously, after the two-way signal transfer, the total effect of ionospheric delay could be further reduced. However, to better eliminate the error, we need to know the $\Delta e = e_{As} - e_{Bs}$. Usually, the global ionospheric TEC map provided by the IGS (International GNSS Service) every 2 hours can be used to eliminate or reduce the delay error. Taking the typical value of TEC, it can be estimated that the ionospheric delay is about 0.1 ns.

There is an error caused by the distances between the two stations and the satellite. The distance between the two stations and satellite is different, which means the signal arrival time from A to S and B to S is different. There is the problem that, when the signal of one station arrives at the satellite, another signal is on the way. Because the geostationary satellite is not strictly relative static with the Earth, the satellite's location is different between the first-arriving and the late-arriving. It will cause the signal path from A to B and B to A to be asymmetrical, seriously influencing the advantages of symmetrical paths of TWSTFT in eliminating errors. Approximately, when the distance converted from the longitude difference between the two ground stations and the satellite longitude is 300 km, the maximum error is about 30 ps. To reduce signal arrival time difference, we can slightly adjust the transmission delay to make them arrive the satellite simultaneously. If the time difference between the two signals arriving at the satellite is less than 5 ms, the effect in TWSTFT will be less than 1 ps [43].

*4.3 Sagnac effect error*

When we transmitted the signal from the station to the satellite, the signal's route is not time-varying [45]. Since the satellite and ground stations are moving, this motion causes the signal-propagation change; the corresponding influence is referred to as the Sagnac effect [46]. After the two-way link difference, the error caused by the Sagnac effect is about 100-200 ns and needs further correction. The Sagnac effect correction for a one-way route from the ground station to satellite is given in a model that provides sufficient accuracy by the following expression [42, 45, 46]:

$$\tau_{sd}^k = \frac{\Omega}{c^2} R\{acos[\tan^{-1}(\tan\varphi^k - f\tan\varphi^k)] + H^k \cos\varphi^k\}\sin(\lambda^k - \lambda^s) \tag{9}$$

where $\Omega$ is the Earth's rotation rate, R is the distance from the satellite to the geocenter, $a$ is Earth's equatorial average radius, $f$ is the flattening of the Earth ellipsoid, $H^k$ (k=A or B) is the height of the station above the ellipsoid, $\lambda^s$ is the longitude of the satellite, and $\varphi^k$ and $\lambda^k$ are the latitude and longitude of the station, respectively.

After the model correction, only about a 10-100 picosecond error will be left. The reason for this residual is that, for the ground observer, the position of a geostationary satellite is not completely fixed. There will be a slight periodic movement with a daily period around a central point (see Figure 4); this will cause the Sagnac effect to change periodically, and its maximum peak to peak amplitude is hundreds of ps. In our study, the influence magnitude is 10-100 ps. At the current accuracy level, it has almost no effect on our experimental results, but it may be considered if higher accuracy is needed. Various error sources and their magnitudes of influence are listed in Table 1.

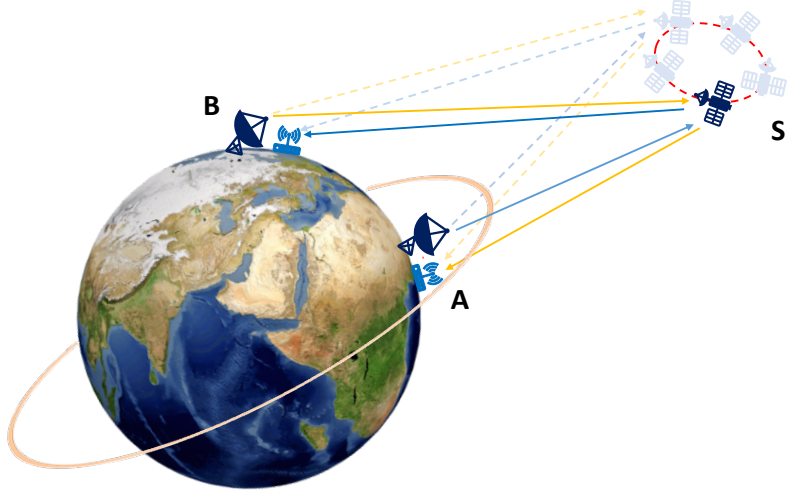

**Figure 4**. The influence of small periodic movement of satellite on TWSTFT. In terrestrial reference frame, the position of satellite relative to station changes over time. That means the $\lambda^s$ and R in Eq. (9) will change over time. This leads to Sagnac effect error, which is no longer a constant but varies with the period of the satellite.

Table 1. Influences of various errors on TWSTFT.

| Error sources | Error magnitude/ps (two-way) | Correction model | Residual error/ps |
|---|---|---|---|
| Time interval counter delay | 10~100 | Zero-baseline calibration | 5 |
| Modems delay | 100 | Zero-baseline calibration | 10 |
| Satellite transparent transponder delay | 80 | Zero-baseline calibration | 10 |
| Transmission and receiving system delay | 200~500 | Zero-baseline calibration | 30 |
| Propagation path geometry delay | < 10 | Neglected | < 10 |
| Tropospheric delay | 10 | Neglected | 10 |
| Ionospheric delay | 100 | Model correction | <10 |
| Asymmetry of station and satellite position delay | 30 | Delay transmission compensation | <1 |
| Sagnac effect delay | $1$~$2\times10^5$ | Model correction | 10~100 |

## 5 Experiments and data processing

*5.1 Experiments*

To verify the feasibility of this method, we need the clocks with high stability to maintain a stable frequency standard and a reliable time-transfer technique to measure the time difference between two positions. Therefore, we used two hydrogen atomic clocks In the experiment; a non-transportable clock $C_A$(H-MASTER VCH-1003A) and a transportable clock $C_B$ (H-MASTER BM2101-01); both of their relative nominal frequency stabilities are $5\times10^{-15}$ in one day. The TWSTFT was used as the time-transfer technique, one of the most accurate time-transfer methods than other GNSS-related techniques [47]. Time transfer by satellite does not have higher stability than fiber link, but it has better performance for long-distance time transfer.

We conducted the experiment in a building at the BIRMM, Beijing. The experiment we conducted was a geopotential difference measurement from 15 December 2016 to 27 December 2016. During this process, $C_B$ was moved to the fifth floor at position B, while $C_A$ was placed on the ground floor (see Fig.5a). The height difference between the fifth floor and ground floor is 22.8 m. After the experiment was conducted, we carried out a zero-baseline measurement from 27 December 2016 to 3 January 2017 to calculate the clock drift and equipment error for calibration. As Fig. 5b shows, clocks $C_A$ and $C_B$ were both placed on the ground floor. The equipment was temperature-stabilized and controlled during the experiment. At every site (A or B), through connection cables, one hydrogen clock, a pulse signal isolation amplifier (PSIA), a modem, and a satellite antenna were connected (see Fig. 5).

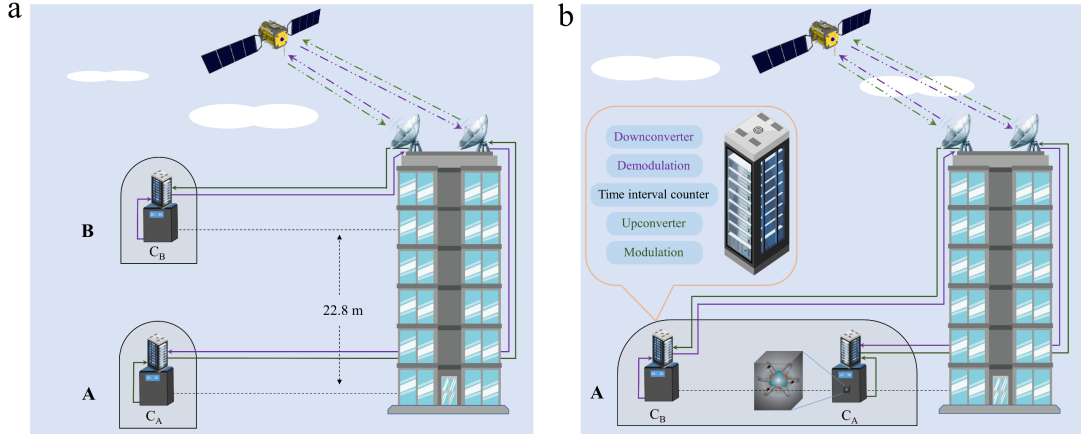

**Figure 5.** Time difference measurement. (**a**) When the geopotential difference measurement was conducted, clock $C_B$ was transferred and placed at B (fifth floor), and clock $C_A$ was still placed at A (ground floor). (**b**) When the zero-baseline measurement was conducted, clocks $C_A$ and $C_B$ were both placed at A.

*5.2 Data processing*

In the experiments, the data we collected were the time difference between clock $C_A$ and $C_B$ corresponding to time. The sampling rate of data is 1 Hz. According to Eq. (1), we just want to calculate $\frac{\Delta t_{AB}}{T}$, which is the slope of data, so it is insignificant whether the two clocks are synchronized in the beginning. What we are concerned with is the running rates of the two clocks.

There are some defects because of an equipment sampling error in the row data (Fig. 6a,b). There are accidental data jumps, outliers, and some missing data. To improve the quality of the data, the following procedures are adopted:

(1) Through the data analysis with a large magnitude of change, we found that the data had some jumps, so we used fitting to restore them to the correct positions to ensure the continuity of all data.

(2) To avoid the influence of outliers on the calculation results, we adopted 3σ criterion (PauTa criterion) to identify and eliminate outliers. The occasional outliers might be due to the fact that during its propagation, the radio frequency signal will suffer from various influences, which causes its distortion. Therefore, in the demodulation process, the sampling decision device cannot accurately reproduce the original 1PPS signal, leading to outliers.

(3) During continuous observation, some missing data may be caused by accidental failure of touch elements in TIC. We used linear interpolation to supplement the missing data.

(4) We used singular spectrum analysis (SSA) [48] to remove periodic terms. This allows better extraction of trend items.

After these procedures, we obtained the 'valid' data. Then, we corrected some errors in time transfer, as discussed in Section 4. In the experiment, we just want to calculate the time interval difference ($\frac{\Delta t_{AB}}{T}$) caused by geopotential difference. In other words, we do not need to consider the synchronization accuracy of time, and we only need to consider the change rate of the time difference between the two sites. Therefore, the errors we need to consider are the changes over time rather than the fixed values over time in equipment delay, ionospheric delay, and the Sagnac effect. Among equipment delays, the time-varying part is the frequency drift of the clock. The output frequency of the atomic clock has a linear drift with time, which can be found in the zero-baseline experiment (Fig. 6c), and could be corrected by the Zero-baseline experiment. In the ionospheric delay correction model, the TEC value changes in real-time. However, because the TEC values on the signal path are all the same (two antennas in one building), the influences on the result can be ignored. The time delay caused by the Sagnac effect is a fixed value because both the variation of the Earth's rotation speed and satellite orbit is minimal, which can be considered as constants under the current experimental accuracy [45]. Hence, the time delay caused by the Sagnac effect will not influence the experimental result at the present accuracy requirement.

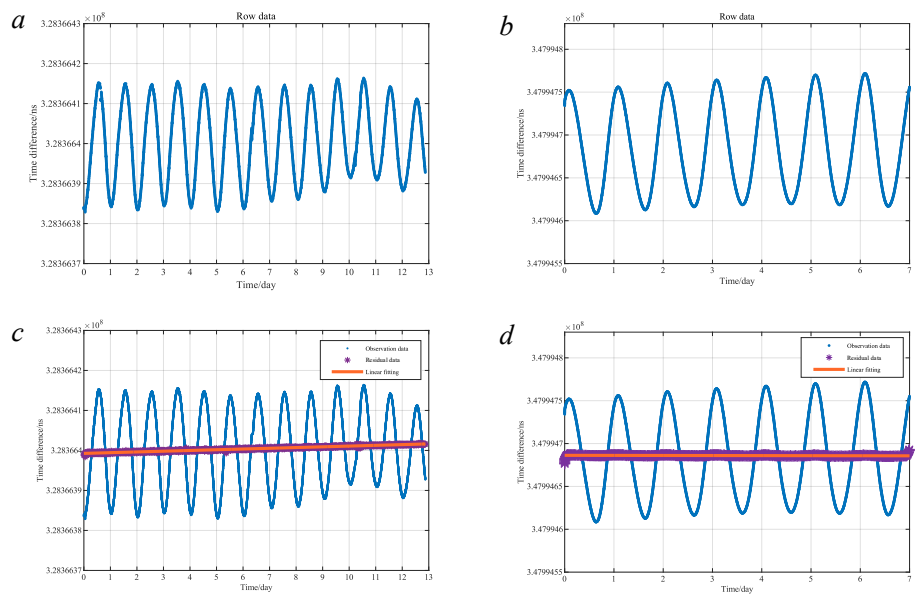

**Figure 6**. Comparison between two clocks. (**a**) Two clocks on two floors; row data from 15 December 2016 to 27 December 2016 for a total of 13 days. (**b**) Two clocks on the same ground floor; row data from 27 December 2016 to 3 January 2017 for a total of 7 days. (**c**) Two clocks on two floors; residual data after processing. (**d**) Two clocks on the same ground floor; residual data after processing. Blue curves denote original observations, purple curves denote the data after removing the diurnal terms from the original data noted as residual data, and orange lines are linear fittings of the purple curves.

Then, we obtained the data with an apparent linear trend and used a linear function to fit the data. As shown in Fig. 6c and d, the purple points show the residual data after removing diurnal signals from the original observation data; the orange line donates the fitting line. In Fig. 6c, the data show relatively intense jumps in the geopotential measurement. The reason for the jumps is that the temperature-control equipment was not particularly stable, which influenced the accuracy of the observations. Nevertheless, on the whole, this is in good agreement with the linear fitting line. This trend mainly includes the frequency drift of the clock, equipment error, and the time interval difference caused by geopotential differences. Therefore, we only need to eliminate the frequency drift of the clock and equipment error by zero-baseline measurement to obtain the results. In Fig. 6d, in the whole zero-baseline comparison observation period, the data are relatively smooth and have a strong trend, which is conducive to the calculation of results.

## 6. Results

After calculation, the slopes of the geopotential comparison experiment and zero-baseline experiment are $k_{geo}$ = 2.11639×10⁻¹⁵ and $k_{zero}$ = -0.93617×10⁻¹⁵, respectively. The slope of the zero-baseline measurement is the constant system shift; therefore, subtracting it from that of the geopotential comparison experiment could determine the difference of the clock running rates between C$_A$ and C$_B$. To calculate the difference of the clock running rates ($\frac{\Delta t_{AB}}{T}$), we differenced the two slopes:

$$\frac{\Delta t_{AB}}{T} = k_{geo} - k_{zero} = 3.0526 \times 10^{-15} \qquad (10)$$

where $k_{remote}$ and $k_{zero}$ denote the slopes of the geopotential comparison experiment and zero-baseline experiment, respectively.

Based on Formulas (1)-(3), the measured height difference is $\Delta H = \frac{\Delta t_{AB}}{T}\frac{c^2}{g} = 28.00m$. The residual standard deviation of the regression equation is expressed as [49]:

$$S = \sqrt{\frac{\sum_{i=1}^{n}(y_i - \hat{y}_i)^2}{n-2}} \qquad (11)$$

where $n$ is the total number of the sampling interval of the whole time series, $y_i$ denotes the $i$th observation, and $\hat{y}_i$ denotes the $i$th fitting value. The uncertainty of the slope is given by [49]

$$u = \frac{S}{\sqrt{\sum_{i=1}^{n}(x_i - \bar{x})^2}} \qquad (12)$$

where S is the residual standard deviation of the regression equation, $x_i$ is the time of the $i$th observation, and $\bar{x}$ is the mean value of $x$.

Based on Formulas (11) and (12), we obtained the uncertainties of the slopes of the geopotential comparison experiment and zero-baseline experiment, written as $u_{geo}$ = 0.26×10⁻¹⁵ and $u_{zero}$ = 0.52×10⁻¹⁵, respectively. On the basis of the propagation law of errors, the uncertainty of the difference of the clock running rates ($\frac{\Delta t_{AB}}{T}$) is

$$u_D = \sqrt{u_{remote}^2 + u_{zero}^2} = 0.58 \times 10^{-15} \qquad (13)$$

Finally, substituting the uncertainty values into Equations (1) and (3), we obtained the accuracy of the measurements, 5.4 m.

The relevant results are shown in Table 2.

Table 2. Results of geopotential comparison and zero-baseline comparison.

|  | Geopotential comparison | Zero-baseline |
| --- | --- | --- |
| Slope | 2.11639×10⁻¹⁵ | -0.93617×10⁻¹⁵ |
| The uncertainty of the slope | 0.26×10⁻¹⁵ | 0.52×10⁻¹⁵ |

| | |
|---|---|
| Measured height difference between A and B (m) | 28.0±5.4 |
| True value (m) | 22.8 |
| Deviation (m) | 5.2±5.4 |

## 7. Discussions and conclusions

The experimental results in period 1 (geopotential comparison measurement) provide an uncertainty of the slope $0.26\times10^{-15}$, which has better accuracy than period 2 (zero-baseline comparison measurement) $0.52\times10^{-15}$. The reason for this result may be that the observation time of period 1 is longer than period 2; a longer observation time is helpful to weaken the influence of observation noise on experimental results.

Based on TWSTFT observations, we determined the height difference between A and B as 28.0±5.4 m. The bias between the measured value and the corresponding true value of 22.8 m is 5.2 m. Concerning the observation precision level of 5.4 m, the results are acceptable. Although the results reached expectations, there are still some problems worth further exploration. We cannot explain why there is strong periodicity in the observed data, but it is conjectured that these are connected with the atomic clock's performances and the relative motion of the satellite. The temperature could significantly influence the running rate of the atomic clock [50, 51]. Due to the limitation of the experiment conditions, the ambient temperature of the atomic clock is not completely constant, and there is a certain fluctuation, which may be the reason for the jump in the observed data. In addition, the change of satellite orbit may also have an impact on the results. As described in Section 4.3, the satellite's orbit with a daily period, which is not fixed and changes every day.

Here, we employed two hydrogen MACs and TWSTFT technology to measure the height difference. The results of the experiment indicate that the accuracy of TWSTFT used in this experiment is great enough. Our experimental results are preliminary. Further studies and experiments are needed to advance this research.

**Author Contributions:** Initiation and experiment design, W.-B.S.; conducting experiments, X.S., C.-C.H., K.-C.W., Z.-Y.S. and W.-B.S.; methodology, P.C. and W.-B.S.; software, P.C.; data processing, P.C.; validation, P.C., X.S. and C.-C.H; formal analysis, P.C., W.-B.S. and K.-C.W.; investigation, P.C. and W.-B.S.; original draft preparation, P.C.; writing-review and editing, P.C., W.-B.S., X.S., C.- C.H., K.-C.W. and Z.-Y.S.; visualization, P.C.; supervision, W.-B.S.; project administration, W.-B.S. All authors have read and agreed to the published version of the manuscript.

**Acknowledgements:** This study is supported by National Natural Science Foundation of China (Grant nos. 41721003, 42030105, 41631072, 41804012, 41874023, 41974034), Space Station Project (2020)228 and Natural Science Foundation of Hubei Province (grant No. 2019CFB611).

**Data Availability Statement:** The data that support the plots within this paper and other findings of this study are available from the corresponding author upon reasonable request.